\documentclass[runningheads]{llncs}
\usepackage[T1]{fontenc}
\usepackage{graphicx}
\usepackage{ upgreek }
\usepackage[font=small,belowskip=-5pt,aboveskip=0pt]{caption}

\usepackage{multirow}
\usepackage{graphicx}
\usepackage{xspace}
\usepackage{amsmath}
\usepackage{amsfonts}
\usepackage[table,xcdraw]{xcolor}
\usepackage{wrapfig,lipsum,booktabs}

\newcommand{\ie}{i.e.,~}

\def\MCD{MC Dropout\@\cite{gal2016dropout}\xspace}
\def\CD{Concrete Dropout\@\cite{gal2017concrete}\xspace}
\def\NE{Ensemble\@\cite{lakshminarayanan2017simple}\xspace}
\def\BE{Batch Ensemble\@\cite{wen2020batchensemble}\xspace}
\def\RBNN{Rank1 BNN\@\cite{dusenberry2020rank1}\xspace}
\def\LPBNN{LP-BNN\@\cite{franchi2020encoding}\xspace}
\def\SWAG{SWAG\@\cite{maddox2019swag}\xspace}
\def\MSWAG{Mulit-SWAG\@\cite{wilson2020multiswag}\xspace}

\begin{document}
%
\title{Benchmarking Scalable Epistemic Uncertainty Quantification in Organ Segmentation}
\titlerunning{Benchmarking Scalable Epistemic UQ in Organ Segmentation}

\author{Jadie Adams\inst{1,2} \and
Shireen Y. Elhabian\inst{1,2}}
%
\authorrunning{Adams and Elhabian}
%
\institute{Scientific Computing and Imaging Institute, University of Utah, UT, USA \and
School of Computing, University of Utah, UT, USA \\
\email{ jadie.adams@utah.edu, shireen@sci.utah.edu }
}

\maketitle  
\begin{abstract}
Deep learning based methods for automatic organ segmentation have shown promise in aiding diagnosis and treatment planning. However, quantifying and understanding the uncertainty associated with model predictions is crucial in critical clinical applications. While many techniques have been proposed for epistemic or model-based uncertainty estimation, it is unclear which method is preferred in the medical image analysis setting. This paper presents a comprehensive benchmarking study that evaluates epistemic uncertainty quantification methods in organ segmentation in terms of accuracy, uncertainty calibration, and scalability. We provide a comprehensive discussion of the strengths, weaknesses, and out-of-distribution detection capabilities of each method as well as recommendations for future improvements. These findings contribute to the development of reliable and robust models that yield accurate segmentations while effectively quantifying epistemic uncertainty. 
\end{abstract}
\setcounter{footnote}{0}
\section{Introduction}
Deep learning systems have made significant strides in automating organ segmentation from 3D medical images. 
Segmentation networks can be efficiently integrated into image processing pipelines, facilitating research and clinical use (\ie tumor segmentation in radiotherapy \cite{savjani2022automated} and hippocampus segmentation for neurological disease analysis \cite{chupin2009fully}). 
However, these systems also introduce new challenges and risks compared to traditional segmentation processes, including issues of bias, errors, and lack of transparency.
Deep networks are prone to providing overconfident estimates and thus cannot be blindly trusted in sensitive decision-making scenarios without the safeguard of granular uncertainty quantification (UQ) \cite{li2023trustworthy,liang2022advances}.
UQ is the process of estimating and representing the uncertainty associated with predictions made by deep neural networks.
UQ provides necessary insight into the reliability and confidence of the model’s predicted segmentation.
In the context of organ segmentation, areas near organ boundaries can be uncertain due to the low contrast between the target organ and surrounding tissues \cite{wang2019aleatoric}. Pixel or voxel-level uncertainty estimates can be used to identify potential incorrect regions or guide user interactions for refinement \cite{wang2019aleatoric,prassni2010uncertainty}. 
This enables quality control of the segmentation process and the detection of out-of-distribution (OOD) samples.

Two forms of uncertainty are distinguished in deep learning frameworks: aleatoric and epistemic\cite{hullermeier2021aleatoric}. 
\textit{Aleatoric uncertainty} refers to the inherent uncertainty in the input data distribution that cannot be reduced \cite{kendall2017uncertainties} (\ie uncertainty resulting from factors like image acquisition noise, over-exposure, occlusion, or a lack of visual features \cite{simpson2019decathlon}).
Aleatoric uncertainty is typically quantified by adjusting the model to be stochastic (predicting a distribution rather than a point-wise estimate \cite{kendall2017uncertainties}) or by  
methods such as test time augmentation \cite{wang2019aleatoric,shanmugam2021better}.
\textit{Epistemic uncertainty} is model-based and arises from a lack of knowledge or uncertainty about the model's parameters due to limited training data or model complexity. 
Capturing epistemic uncertainty is considerably more difficult as it cannot be learned as a function of the input but rather requires fitting a distribution over model parameters. 
Several approaches have been proposed to accomplish this, but many of them significantly increase the computational cost and memory footprint and may impact prediction accuracy \cite{dusenberry2020rank1}.
There is no ubiquitous method for epistemic UQ in segmentation networks, as each proposed technique has its own trade-offs and limitations.

This study benchmarks Bayesian and frequentist epistemic UQ methods for organ segmentation from 3D CT scans in terms of scalability, segmentation accuracy, and uncertainty calibration using multiple datasets. 
While previous benchmarks (e.g., \cite{ng2022cardiacsegbenchmark,sahlsten2023tumorUQ,mehrtash2020confidence}) have been conducted on small subsets of such methods, there is a need for a comprehensive evaluation. To the best of our knowledge, this work provides the most extensive benchmarking of scalable methods for epistemic UQ in medical segmentation.
The key contributions are as follows: 
\begin{enumerate}
    \item We conduct a benchmark evaluation of scalable methods for epistemic UQ in medical image segmentation, including  deep ensemble \cite{rahaman2021ensemble}, batch ensemble \cite{wen2020batchensemble}, Monte Carlo dropout \cite{gal2016dropout}, concrete dropout \cite{gal2017concrete}, Rank-1 Bayesian Neural Net (BNN) \cite{dusenberry2020rank1}, latent posterior BNN \cite{franchi2020encoding}, Stochastic Weight Averaging (SWA) \cite{izmailov2018swa}, SWA Gaussian (SWAG) \cite{maddox2019swag}, and Multi-SWAG \cite{wilson2020multiswag}.
    \item We evaluate these methods in detecting out-of-distribution (OOD) instances, which is an important aspect of robust uncertainty estimation.
    \item We provide a comprehensive discussion of the strengths and weaknesses of the evaluated methods, enabling a better understanding of their performance characteristics and potential improvements.
    \item To facilitate further research and reproducibility, we provide an open-source PyTorch implementation of all benchmarked methods.\footnote{Source code is publicly available: \url{https://github.com/jadie1/MedSegUQ}}
\end{enumerate}

\section{Epistemic Uncertainty Quantification Techniques}

Modeling epistemic uncertainty in a scalable manner poses significant challenges as it entails placing distributions over model weights.
Both Bayesian and frequentist methods have been proposed, with Bayesian approaches aiming to directly estimate the posterior distribution over the model's parameters, while frequentist methods use ensembles of models to approximate the posterior empirically.

In Bayesian deep learning, obtaining an analytical solution for the posterior is often intractable, necessitating the use of approximate posterior inference techniques such as variational inference \cite{VarInference}.
The most common Bayesian technique for UQ is \textbf{Monte Carlo (MC) dropout} sampling, as it provides a fast, scalable solution for approximate variational inference \cite{gal2016dropout}.
In MC dropout, uncertainty is captured by the spread of predictions resulting from sampled dropout masks in inference. 
However, obtaining well-calibrated epistemic UQ with dropout requires a time-consuming grid search to tune layer-wise dropout probabilities.
\textbf{Concrete dropout} \cite{gal2017concrete} was proposed to address this limitation by automatically optimizing layer-wise dropout probabilities along with the network weights. 
Certain Bayesian approaches for approximate inference are excluded from this benchmark due to their limited scalability and tendency to underfit \cite{dusenberry2020rank1,gal2016dropout,lakshminarayanan2017simple}, such as sampling-based Markov chain Monte Carlo \cite{chen2014stochastic}, Bayes by Backprop \cite{blundell2015weight}, and variational inference based methods \cite{VarInference}.
These techniques rely on structured or factorized distributions with tied parameters, have a high computational cost, and are generally slow to converge \cite{gal2016dropout}.
Additionally, previous work has shown that such methods perform similarly to the much more lightweight MC dropout approach in medical image segmentation UQ \cite{ng2018estimating}. 

\textbf{Deep ensembles} are an effective and popular frequentist method for UQ \cite{hu2019mbpep}. 
Ensembling involves training multiple independent networks (or ensemble members) with different initialization then aggregating predictions for improved robustness \cite{fort2019deep}. 
The spread or variability among the predictions of the ensemble members effectively captures the epistemic uncertainty \cite{lakshminarayanan2017simple}. 
It has been shown that deep ensemble models can provide a better approximation than standard Bayesian methods \cite{wilson2020multiswag}. 
The main drawback of deep ensembles lies in their computational and memory costs, which increase linearly with the number of ensemble members.
To address the trade-off between accuracy and scalability, \textbf{batch ensemble} \cite{wen2020batchensemble} has been proposed. 
Batch ensemble \cite{wen2020batchensemble} compromises between a single network and an ensemble by employing shared weight matrices and lightweight rank-1 ensemble members.
The concept of batch ensembling has also been applied to improve the scalability of Bayesian Neural Networks (BNNs). \textbf{Rank-1 BNN} \cite{dusenberry2020rank1} reduces computational complexity by utilizing a rank-1 parameterization in variational inference.
\textbf{Latent Posterior BNN (LP-BNN)} \cite{franchi2020encoding} was proposed to improve scalability further by learning the posterior distribution of lower-dimensional latent variables derived from rank-1 vectors. 
This is accomplished by training layer-wise variational autoencoders (VAEs) \cite{kingma2014autoencoding} on the rank-1 vectors. The latent space of these VAEs can then be sampled, providing a distribution of rank-1 weights.

Additional methods of epistemic UQ have been developed based on the stochastic weight averaging (SWA) technique \cite{izmailov2018swa}.
SWA was proposed to enhance generalization in deep learning by estimating the mean of the stationary distribution of SGD iterates.
In SWA, final model weights are defined by averaging the weights traversed during SGD after initial convergence.
\textbf{SWA-Gaussian (SWAG)} \cite{maddox2019swag}
fits a Gaussian distribution to the traversed weights, creating an approximate posterior distribution over the weights that can be sampled for Bayesian model averaging. 
It has been shown that combining traditional Bayesian methods with ensembling improves the fidelity of approximate inference via multimodal marginalization, resulting in a more robust, accurate model \cite{wilson2020multiswag}. 
Based on this observation, \textbf{Multi-SWAG} \cite{wilson2020multiswag} was proposed as an ensemble of SWAG models.
These approaches offer alternative ways to capture epistemic uncertainty by leveraging the ensemble characteristics and combining them with Bayesian principles. They aim to improve the fidelity of approximate inference and provide scalable solutions for epistemic UQ in deep learning tasks.

\section{Experimental Design}

We utilize the residual U-Net architecture originally proposed for cardiac left-ventricle segmentation \cite{kerfoot2019rUnet} as a base architecture to compare the epistemic UQ techniques. 
This model is comprised of residual units of 3D convolutional layers with batch normalization and PReLU activation.
As a baseline for UQ calibration analysis, we consider the predicted segmentation probabilities. Specifically, for the \textbf{base} model, we quantify voxel-wise UQ as: $UQ = 1 - C$, where confidence, $C$, is the maximum of the foreground and background softmax probabilities. This is not strictly a measure of epistemic UQ, as there is no notion of posterior approximation and marginalization. However, this formulation of UQ has been shown to correlate with prediction error and is useful for OOD detection \cite{hendrycks2016baseline} and thus provides an evaluation baseline. 

We benchmark the following aforementioned scalable methods for epistemic UQ: \NE, \BE, \MCD, \CD, \RBNN, \LPBNN, \SWAG, and \MSWAG.
In implementing dropout and rank1-based methods, dropout and batch ensemble are applied to every convolutional layer respectively. 
Models are trained on $(96 \times 96 \times 96)$ patches of the input images scaled to an intensity of $[0,1]$. 
A validation set is used to assess the convergence of each model. Models are trained until the performance on the validation set has not improved in 100 epochs. The model weights resulting from the epoch with the best validation performance are used in the evaluation of held-out testing data. 
Implementation and tuned hyperparameter values for each model are provided in the GitHub repository. 

\subsection{Datasets}
We utilize two open-source datasets from the Medical Segmentation Decathlon \cite{simpson2019decathlon} in evaluation: the spleen and pancreas. These datasets comprise of 3D CT images and corresponding manual ground truth segmentations. 
The \textbf{spleen} dataset was selected to provide a typical medical image analysis scenario where data is scarce and varied. There are only 41 instances in the spleen dataset, and the size of the spleen versus the background varies widely.
The \textbf{pancreas} dataset was selected to evaluate OOD detection accuracy. This dataset contains cancerous cases with segmentations of both the pancreas organ and tumor masses.
We analyze the accuracy of joint segmentation of the pancreas and tumors, holding out the cases with the largest tumors an OOD test set. 

\subsection{Metrics}
In 3D image segmentation, accuracy is typically assessed via the \textbf{Dice Similarity Coefficient (DSC)} between manual annotations and the model’s prediction. The DSC metric captures the percentage of voxel-wise agreement between a predicted segmentation and its corresponding ground truth.
In these experiments the target organs are small compared to the total image size. Because of this, we only include the foreground in DSC calculations so that the DSC value is not overwhelmed by the background signal. 

To measure overall uncertainty calibration, we consider the correlation between estimated epistemic uncertainty and prediction error (100 - DSC). We report the \textbf{Pearson correlation coefficient (r)} between error and sum of the uncertainty map, where a higher r-value indicates better correlation. 

Finally, we assess segmentation accuracy and uncertainty quality jointly via \textbf{error-retention curves} \cite{lakshminarayanan2017simple,malinin2022shifts}. Error-retention curves depict a given error metric over a dataset as ground-truth labels replace a model’s predictions in order of decreasing estimated uncertainty. The \textbf{area under the curve (AUC)} is reduced as the overall error is decreased, as well as the correlation between error and uncertainty is increased. We report the \textbf{area under the error retention curve (R-AUC)} using 100 - DSC as the error metric. Smaller R-AUC indicates both better segmentation accuracy and uncertainty calibration.

\section{Results}

\subsection{Spleen}
\begin{figure}[ht!]
    \centering
    \includegraphics[width=\textwidth]{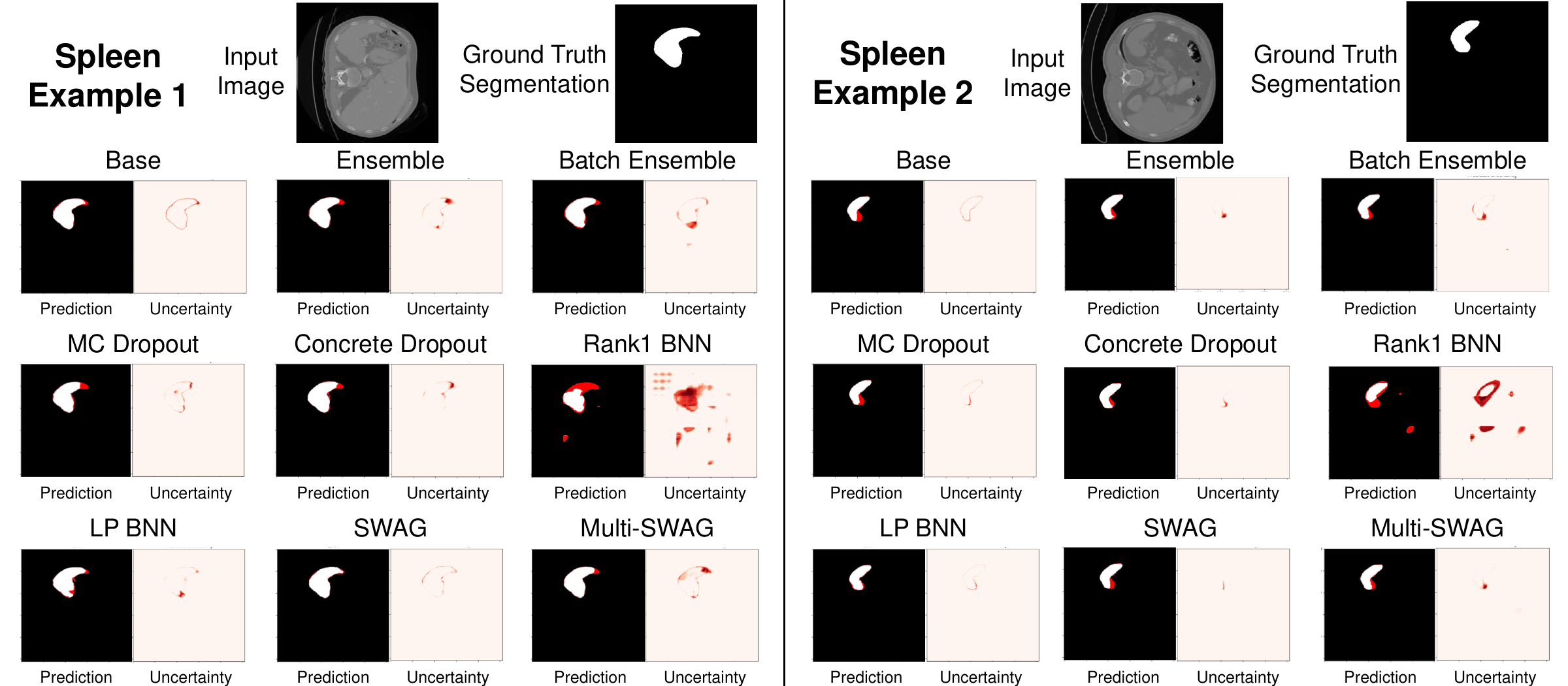}
    \caption{Slices of two spleen examples are provided with the segmentation resulting from each model with error overlayed in red. Additionally, predicted uncertainty maps are shown where darker red indicates higher uncertainty.} \label{fig:spleen}
\end{figure}

Because the spleen dataset is comprised of only 41 examples, we employ K-folds to define the training, validation, and test sets. The data is split into 70\% train, 10\% validation, and 20\% held-out test using five different folds. For each fold, a separate model of every form is trained. 
In this manner, results are reported across the entire dataset, where the predicted segmentation of each image is acquired from the model for which the image was held out.
The results are reported in Table \ref{table:spleen}, and qualitative visualizations are provided in Fig. \ref{fig:spleen}.
For Ensemble and Batch Ensemble models four ensemble members are used, thus four predictions are averaged and used for UQ estimation.
\begin{wraptable}{r}{7.5cm}
\vspace{-.1in}
\caption{Spleen results: Mean and standard deviation values across held-out data, best values in bold.}
\label{table:spleen}
\centering
\resizebox{.62\textwidth}{!}{%
\begin{tabular}{ccccccc}
\cline{2-4}
\multicolumn{1}{c|}{} & \multicolumn{1}{c}{DSC $\uparrow$} & \multicolumn{1}{c}{r $\uparrow$} & \multicolumn{1}{c|}{R-AUC $\downarrow$} &  &  &  \\ \cline{1-4}
\multicolumn{1}{|l|}{Base} & 89.15$\pm$9.55 & 0.56 & \multicolumn{1}{c|}{2.54$\pm$4.52} &  &  &  \\
\cline{1-4}
 &  &  &  &  &  &  \\ \hline
\multicolumn{1}{|c|}{} & \multicolumn{3}{c|}{4 Samples} & \multicolumn{3}{c|}{30 Samples} \\
\multicolumn{1}{|c|}{\multirow{-2}{*}{\textbf{Method}}} & \multicolumn{1}{c}{DSC $\uparrow$} & \multicolumn{1}{c}{r $\uparrow$} & \multicolumn{1}{c|}{R-AUC $\downarrow$} & \multicolumn{1}{c}{DSC $\uparrow$} & \multicolumn{1}{c}{r $\uparrow$} & \multicolumn{1}{c|}{R-AUC $\downarrow$} \\ \hline
\multicolumn{1}{|l|}{\NE} & \textbf{92.77}$\pm$5.34 & 0.37 & \multicolumn{1}{c|}{0.71$\pm$1.09} & N/A & N/A & \multicolumn{1}{c|}{N/A} \\
\multicolumn{1}{|l|}{\BE} & 86.87$\pm$14.21 & 0.40 & \multicolumn{1}{c|}{0.58$\pm$1.09} & N/A & N/A & \multicolumn{1}{c|}{N/A} \\
\multicolumn{1}{|l|}{\MCD} & 86.16$\pm$19.07 & -0.07 & \multicolumn{1}{c|}{1.54$\pm$4.06} & 86.40$\pm$18.33 & -0.08 & \multicolumn{1}{c|}{1.31$\pm$3.79} \\
\multicolumn{1}{|l|}{\CD} & 90.21$\pm$6.42 & 0.45 & \multicolumn{1}{c|}{\textbf{0.42$\pm$0.47}} & 90.23$\pm$6.49 & 0.59 & \multicolumn{1}{c|}{\textbf{0.28$\pm$0.35}} \\
%
\multicolumn{1}{|l|}{\RBNN} & 55.01$\pm$19.17 & 0.10 & \multicolumn{1}{c|}{3.69$\pm$3.47} & 66.36$\pm$16.84 & 0.11 & \multicolumn{1}{c|}{0.79$\pm$1.45} \\
\multicolumn{1}{|l|}{\LPBNN} & 87.77$\pm$10.66 & 0.57 & \multicolumn{1}{c|}{0.98$\pm$1.27} & 87.74$\pm$10.83 & 0.47 & \multicolumn{1}{c|}{0.93$\pm$1.48} \\
\multicolumn{1}{|l|}{\SWAG} & 87.80$\pm$15.53 & 0.18 & \multicolumn{1}{c|}{0.99$\pm$2.47} & 87.80$\pm$15.69 & 0.20 & \multicolumn{1}{c|}{0.74$\pm$1.71} \\
\multicolumn{1}{|l|}{\MSWAG} & 92.49$\pm$10.18 & \textbf{0.69} & \multicolumn{1}{c|}{0.69$\pm$0.84} & \textbf{93.11$\pm$10.07} & \textbf{0.64} & \multicolumn{1}{c|}{0.57$\pm$0.77} \\ \hline
 &  &  &  &  &  &  \\
 &  &  &  &  &  & 
\end{tabular}%
}
\vspace{-.5in}
\end{wraptable}

For methods that fit a distribution over weights, any number of weight samples can be used to provide an average prediction and UQ estimation. 
Additional samples improve accuracy and calibration but also increase inference time. In evaluating such methods, we elect to use 4 and 30 samples for a comprehensive comparison. 

\subsection{Pancreas}
\begin{figure}[htbp]
\centering
  \begin{minipage}{0.3\textwidth}
    \centering
    \includegraphics[width=.7\linewidth]{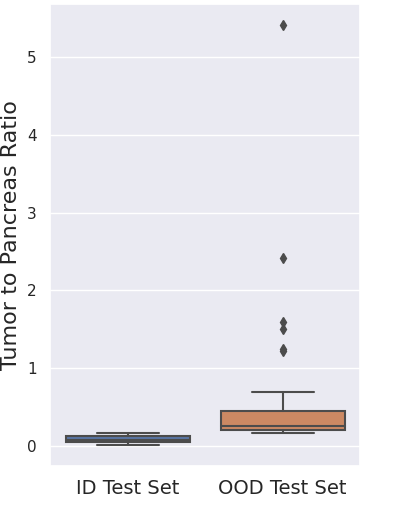}
    \caption{Pancreas test set ratio box plots.}
    \label{fig:ratios}
  \end{minipage}%
  \begin{minipage}{0.7\textwidth}
    \centering
    \captionsetup{type=table}
    \caption{Pearson correlation coefficients across both pancreas test sets are reported, where error = 100 - DSC.}
    \label{table:ratios}
    \resizebox{\textwidth}{!}{%
        \begin{tabular}{l|l|ccc|l}
        \cline{2-5}
         & \multicolumn{1}{c|}{\multirow{2}{*}{Model}} & \multicolumn{3}{c|}{r values} &  \\
         & \multicolumn{1}{c|}{} & Ratio/Error & UQ/Error $\uparrow$ & UQ/Ratio $\uparrow$ &  \\ \cline{2-5}
         & \multicolumn{1}{l|}{Base} & 0.42 & -0.08 & -0.06 &  \\
         \hspace{.7in} & \NE & 0.44 & 0.22 & -0.01 & \hspace{.7in} \\
         & \BE & 0.34 & 0.13 & 0.10 &  \\
         & \MCD & 0.46 & 0.35 & 0.10 &  \\
         & \CD & 0.48 & 0.28 & 0.08 &  \\
         & \RBNN & 0.02 & -0.32 & 0.06 &  \\
         & \LPBNN & 0.38 & 0.2 & 0.06 &  \\
         & \SWAG & 0.34 & \textbf{0.46} & \textbf{0.11} &  \\
         & \MSWAG & 0.39 & 0.22 & -0.04 &  \\ \cline{2-5}
        \end{tabular}%
    }
  \end{minipage}
\end{figure}

The pancreas dataset is used to analyze the robustness and uncertainty calibration of the various methods in the case of OOD examples. 
To this end, we calculate the ratio of tumor to pancreas voxels in the ground truth segmentations. 
We hold out the 50 instances with the largest tumor ratio as a test set. This provides an OOD test set, as the models are trained only on examples with smaller tumors. 
The remaining 231 image/segmentation pairs are randomly split into a single training, validation, and in-distribution (ID) test set using a 70\%, 10\%, 20\% split. 
Fig. \ref{fig:ratios} displays the distributions of tumor-to-pancreas ratios in the ID and OOD test sets. 
The results are reported on both the ID and OOD test set in Table \ref{table:pancreas}.
Additionally, the correlation between the tumor-to-pancreas ratio and the estimated uncertainty across both test sets is reported in Table \ref{table:ratios}. 
We expect well-calibrated UQ to correlate with the tumor-to-pancreas ratio as the models are not exposed examples with large tumors in training. However, none of the epistemic UQ quantification methods proved a strong correlation with the ratio, suggesting these models are not effective in accurate OOD detection.

\vspace{-.1in}
\begin{table}[ht]
\caption{Pancreas results: Mean and standard deviation values across both held-out test sets are reported with the best values marked in bold.}
\label{table:pancreas}
\resizebox{\textwidth}{!}{%
\begin{tabular}{l
>{\columncolor[HTML]{DAE8FC}}l 
>{\columncolor[HTML]{DAE8FC}}l 
>{\columncolor[HTML]{DAE8FC}}l 
>{\columncolor[HTML]{DAE8FC}}l 
>{\columncolor[HTML]{DAE8FC}}l 
>{\columncolor[HTML]{DAE8FC}}l 
>{\columncolor[HTML]{FFCE93}}l 
>{\columncolor[HTML]{FFCE93}}l 
>{\columncolor[HTML]{FFCE93}}l 
>{\columncolor[HTML]{FFCE93}}l 
>{\columncolor[HTML]{FFCE93}}l 
>{\columncolor[HTML]{FFCE93}}l }
\multicolumn{1}{c}{} & \multicolumn{6}{c}{\cellcolor[HTML]{DAE8FC}\textbf{In-Distribution Test Set}} & \multicolumn{6}{c}{\cellcolor[HTML]{FFCE93}\textbf{Out-of-Distribution Test Set}} \\ \cline{5-7} \cline{11-13} 
\multicolumn{1}{c}{} & \multicolumn{1}{c}{\cellcolor[HTML]{DAE8FC}} & \multicolumn{1}{c}{\cellcolor[HTML]{DAE8FC}} & \multicolumn{1}{c|}{\cellcolor[HTML]{DAE8FC}} & \multicolumn{1}{c}{\cellcolor[HTML]{DAE8FC}DSC} & \multicolumn{1}{c}{\cellcolor[HTML]{DAE8FC}R} & \multicolumn{1}{c|}{\cellcolor[HTML]{DAE8FC}R-AUC} & \multicolumn{1}{c}{\cellcolor[HTML]{FFCE93}} & \multicolumn{1}{c}{\cellcolor[HTML]{FFCE93}} & \multicolumn{1}{c|}{\cellcolor[HTML]{FFCE93}} & \multicolumn{1}{c}{\cellcolor[HTML]{FFCE93}DSC} & \multicolumn{1}{c}{\cellcolor[HTML]{FFCE93}R} & \multicolumn{1}{c|}{\cellcolor[HTML]{FFCE93}R-AUC} \\ \cline{4-7} \cline{10-13} 
\multicolumn{1}{c}{} & \multicolumn{1}{c}{\cellcolor[HTML]{DAE8FC}} & \multicolumn{1}{c|}{\cellcolor[HTML]{DAE8FC}} & \multicolumn{1}{l|}{\cellcolor[HTML]{DAE8FC}Base} & \cellcolor[HTML]{DAE8FC}71.02$\pm$14.49 & \cellcolor[HTML]{DAE8FC}0.12 & \multicolumn{1}{l|}{\cellcolor[HTML]{DAE8FC}8.66$\pm$7.36} & \multicolumn{1}{c}{\cellcolor[HTML]{FFCE93}} & \multicolumn{1}{c|}{\cellcolor[HTML]{FFCE93}} & \multicolumn{1}{l|}{\cellcolor[HTML]{FFCE93}Base} & \cellcolor[HTML]{FFCE93}67.32$\pm$18.35 & \cellcolor[HTML]{FFCE93}0.04 & \multicolumn{1}{l|}{\cellcolor[HTML]{FFCE93}8.95$\pm$7.03} \\ \cline{4-7} \cline{10-13} 
\multicolumn{1}{c}{} & \multicolumn{1}{c}{\cellcolor[HTML]{DAE8FC}} & \multicolumn{1}{c}{\cellcolor[HTML]{DAE8FC}} & \multicolumn{1}{c}{\cellcolor[HTML]{DAE8FC}} & \multicolumn{1}{c}{\cellcolor[HTML]{DAE8FC}} & \multicolumn{1}{c}{\cellcolor[HTML]{DAE8FC}} & \multicolumn{1}{c}{\cellcolor[HTML]{DAE8FC}} & \multicolumn{1}{c}{\cellcolor[HTML]{FFCE93}} & \multicolumn{1}{c}{\cellcolor[HTML]{FFCE93}} & \multicolumn{1}{c}{\cellcolor[HTML]{FFCE93}} & \multicolumn{1}{c}{\cellcolor[HTML]{FFCE93}} & \multicolumn{1}{c}{\cellcolor[HTML]{FFCE93}} & \multicolumn{1}{c}{\cellcolor[HTML]{FFCE93}} \\ \hline
\multicolumn{1}{|c|}{} & \multicolumn{3}{c|}{\cellcolor[HTML]{DAE8FC}4 Samples} & \multicolumn{3}{c|}{\cellcolor[HTML]{DAE8FC}30 Samples} & \multicolumn{3}{c|}{\cellcolor[HTML]{FFCE93}4 Samples} & \multicolumn{3}{c|}{\cellcolor[HTML]{FFCE93}30 Samples} \\
\multicolumn{1}{|c|}{\multirow{-2}{*}{Model}} & \multicolumn{1}{c}{\cellcolor[HTML]{DAE8FC}DSC} & \multicolumn{1}{c}{\cellcolor[HTML]{DAE8FC}r} & \multicolumn{1}{c|}{\cellcolor[HTML]{DAE8FC}R-AUC} & \multicolumn{1}{c}{\cellcolor[HTML]{DAE8FC}DSC} & \multicolumn{1}{c}{\cellcolor[HTML]{DAE8FC}r} & \multicolumn{1}{c|}{\cellcolor[HTML]{DAE8FC}R-AUC} & \multicolumn{1}{c}{\cellcolor[HTML]{FFCE93}DSC} & \multicolumn{1}{c}{\cellcolor[HTML]{FFCE93}r} & \multicolumn{1}{c|}{\cellcolor[HTML]{FFCE93}R-AUC} & \multicolumn{1}{c}{\cellcolor[HTML]{FFCE93}DSC} & \multicolumn{1}{c}{\cellcolor[HTML]{FFCE93}r} & \multicolumn{1}{c|}{\cellcolor[HTML]{FFCE93}R-AUC} \\ \hline
\multicolumn{1}{|l|}{\NE} & 72.44$\pm$13.54 & 0.04 & \multicolumn{1}{l|}{\cellcolor[HTML]{DAE8FC}4.70$\pm$3.57} & \multicolumn{1}{c}{\cellcolor[HTML]{DAE8FC}N/A} & \multicolumn{1}{c}{\cellcolor[HTML]{DAE8FC}N/A} & \multicolumn{1}{c|}{\cellcolor[HTML]{DAE8FC}N/A} & \textbf{67.54$\pm$19.58} & 0.15 & \multicolumn{1}{l|}{\cellcolor[HTML]{FFCE93}5.65$\pm$5.64} & \multicolumn{1}{c}{\cellcolor[HTML]{FFCE93}N/A} & \multicolumn{1}{c}{\cellcolor[HTML]{FFCE93}N/A} & \multicolumn{1}{c|}{\cellcolor[HTML]{FFCE93}N/A} \\
\multicolumn{1}{|l|}{\BE} & 64.04$\pm$16.49 & 0.24 & \multicolumn{1}{l|}{\cellcolor[HTML]{DAE8FC}6.30$\pm$4.38} & \multicolumn{1}{c}{\cellcolor[HTML]{DAE8FC}N/A} & \multicolumn{1}{c}{\cellcolor[HTML]{DAE8FC}N/A} & \multicolumn{1}{c|}{\cellcolor[HTML]{DAE8FC}N/A} & 57.08$\pm$21.51 & 0.12 & \multicolumn{1}{l|}{\cellcolor[HTML]{FFCE93}9.77$\pm$10.02} & \multicolumn{1}{c}{\cellcolor[HTML]{FFCE93}N/A} & \multicolumn{1}{c}{\cellcolor[HTML]{FFCE93}N/A} & \multicolumn{1}{c|}{\cellcolor[HTML]{FFCE93}N/A} \\
\multicolumn{1}{|l|}{\MCD} & 70.16$\pm$13.96 & 0.36 & \multicolumn{1}{l|}{\cellcolor[HTML]{DAE8FC}9.97$\pm$7.43} & 70.22$\pm$13.93 & 0.38 & \multicolumn{1}{l|}{\cellcolor[HTML]{DAE8FC}8.90$\pm$6.54} & 67.24$\pm$18.13 & 0.30 & \multicolumn{1}{l|}{\cellcolor[HTML]{FFCE93}10.63$\pm$9.33} & 67.35$\pm$18.20 & 0.33 & \multicolumn{1}{l|}{\cellcolor[HTML]{FFCE93}9.45$\pm$9.18} \\
\multicolumn{1}{|l|}{\CD} & \textbf{73.07$\pm$11.91} & 0.32 & \multicolumn{1}{l|}{\cellcolor[HTML]{DAE8FC}\textbf{4.67$\pm$3.13}} & \textbf{73.06$\pm$11.98} & 0.33 & \multicolumn{1}{l|}{\cellcolor[HTML]{DAE8FC}\textbf{4.29$\pm$2.64}} & 66.99$\pm$18.40 & 0.08 & \multicolumn{1}{l|}{\cellcolor[HTML]{FFCE93}8.39$\pm$9.17} & \textbf{67.83$\pm$18.62} & 0.27 & \multicolumn{1}{l|}{\cellcolor[HTML]{FFCE93}7.54$\pm$8.71} \\
\multicolumn{1}{|l|}{\RBNN} & 12.77$\pm$11.81 & -0.23 & \multicolumn{1}{l|}{\cellcolor[HTML]{DAE8FC}11.37$\pm$8.99} & 17.76$\pm$13.60 & -0.2 & \multicolumn{1}{l|}{\cellcolor[HTML]{DAE8FC}9.07$\pm$8.91} & 9.20$\pm$10.17 & -0.33 & \multicolumn{1}{l|}{\cellcolor[HTML]{FFCE93}12.91$\pm$10.17} & 12.39$\pm$11.85 & -0.41 & \multicolumn{1}{l|}{\cellcolor[HTML]{FFCE93}9.21$\pm$4.72} \\
\multicolumn{1}{|l|}{\LPBNN} & 65.28$\pm$16.18 & 0.36 & \multicolumn{1}{l|}{\cellcolor[HTML]{DAE8FC}4.92$\pm$3.05} & 65.31$\pm$16.10 & 0.21 & \multicolumn{1}{l|}{\cellcolor[HTML]{DAE8FC}4.66$\pm$2.94} & 60.39$\pm$19.34 & 0.14 & \multicolumn{1}{l|}{\cellcolor[HTML]{FFCE93}6.34$\pm$5.03} & 60.16$\pm$19.78 & 0.19 & \multicolumn{1}{l|}{\cellcolor[HTML]{FFCE93}7.26$\pm$6.95} \\
\multicolumn{1}{|l|}{\SWAG} & 66.68$\pm$17.27 & 0.36 & \multicolumn{1}{l|}{\cellcolor[HTML]{DAE8FC}9.32$\pm$8.24} & 66.69$\pm$17.29 & \textbf{0.49} & \multicolumn{1}{l|}{\cellcolor[HTML]{DAE8FC}9.34$\pm$8.14} & 63.20$\pm$20.90 & \textbf{0.31} & \multicolumn{1}{l|}{\cellcolor[HTML]{FFCE93}10.06$\pm$8.16} & 63.07$\pm$21.04 & \textbf{0.45} & \multicolumn{1}{l|}{\cellcolor[HTML]{FFCE93}9.89$\pm$7.86} \\
\multicolumn{1}{|l|}{\MSWAG} & 69.67$\pm$15.06 & \textbf{0.39} & \multicolumn{1}{l|}{\cellcolor[HTML]{DAE8FC}5.57$\pm$4.32} & 69.31$\pm$15.19 & 0.41 & \multicolumn{1}{l|}{\cellcolor[HTML]{DAE8FC}5.46$\pm$4.24} & 64.94$\pm$21.13 & 0.12 & \multicolumn{1}{l|}{\cellcolor[HTML]{FFCE93}\textbf{6.19$\pm$5.86}} & 64.65$\pm$21.28 & 0.14 & \multicolumn{1}{l|}{\cellcolor[HTML]{FFCE93}\textbf{6.01$\pm$5.71}} \\ \hline
\end{tabular}%
}
\end{table}

\subsection{Scalability Comparison}
Table \ref{table:scal} reports the average time and memory requirements associated with training and testing each model on the pancreas dataset. Note that for models that can be sampled,the reported inference time is for a single sample. Inference time scales linearly with the number of samples.
Additionally, note "params size" refers to the memory required to store the parameters and "pass size" refers to the memory required to perform a forward/backward pass through the network. 
\vspace{-.3in}
\begin{table}[ht]
\caption{Scalability comparison: Time reported in seconds and memory size in MB, best values in bold.}
\label{table:scal}
\centering
\resizebox{.9\textwidth}{!}{%
\begin{tabular}{|l|cccccc|}
\hline
Model & Train epochs & Train time & Inference time & Total  params & Params size & Pass size \\ \hline
Base & 222 & 17261 & 0.2587 & \textbf{4808917} & \textbf{19.24} & \textbf{1211.20} \\ 
\NE & 888 & 69045 & 1.0348 & 19235668 & 76.96 & 4844.80 \\
\BE & 341 & 45409. & 0.8698 & 4824513 & \textbf{19.24} & 4842.81 \\
\MCD & \textbf{197} & \textbf{16841} & \textbf{0.2548} & \textbf{4808917} & \textbf{19.24} & \textbf{1211.20} \\
\CD \hspace{.2in} & 259 & 36847 & 0.3694 & 4808934 & \textbf{19.24} & \textbf{1211.20} \\
\RBNN & 1142 & 157100 & 0.7712 & 4835229 & \textbf{19.24} & 4844.81 \\
\LPBNN & 881 & 121343 & 0.8742 & 4957940 & 19.77 & 4844.92 \\
\SWAG & 422 & 31615 & 0.2921 & 9617834 & 38.48 & \textbf{1211.20} \\
\MSWAG & 1688 & 126460 & 1.1684 & 38471336 & 153.92 & 4844.80 \\ \hline
\end{tabular}%
}
\end{table}
\vspace{-.2in}

\section{Discussion and Conclusion}
\vspace{-0.05in}
We conducted a benchmark of scalable epistemic UQ techniques on two challenging organ segmentation tasks. The spleen experiment represented a low training budget scenario, while the pancreas experiment involved significant variation in the shape and size of the organ and tumor masses.
These challenging scenarios, where the base model provides low prediction accuracy, served as stress tests for UQ evaluation. 
We discuss the performance of each model as follows:

\textbf{Base model}: 
While the base models provided competitive accuracy and r values (suggesting instance-level UQ/error correlation) the R-AUC values were low. This indicates the voxel-level UQ did not correlate well with error and thus could not be used to accurately identify erroneous regions. The base model UQ correlates more with the organ boundary than error, as can be seen in Fig. \ref{fig:spleen}.

\textbf{\NE}: As expected, the ensemble model provided an accuracy improvement over the base model. However, it came at the expense of scalability, which is an important consideration in practical applications. More scalable methods outperformed ensembling in terms of R-AUC in these experiments.

\textbf{\BE}: While batch ensemble reduces the memory cost associated with ensembling, it did not provide the same accuracy improvement. This is likely because joint training of the rank1 members proved difficult on these challenging tasks. However, it still provided improved UQ over the base model. 

\textbf{\MCD}: This approach is appealing as it does not increase memory costs or impact the training objective. However, it did not perform as well as concrete dropout, highlighting the importance of tuning layer-wise dropout probabilities. For MC dropout models, we used a dropout rate of 0.1 for all layers. The concrete dropout optimization found a dropout probability of around 0.08 for shallow layers, increasing to around 0.16 for the deepest layer. 

\textbf{\CD}: This technique arguably performed best overall and is desirable as it is scalable and only requires the addition of concrete dropout layers and a loss regularization term. 

\textbf{\RBNN}: This model did not perform well on either task, especially the pancreas segmentation. While the rank1 parameterization greatly improves the scalability of the BNNs, it does not appear to solve the issue of poor convergence that BNNs are prone to suffer from.

\textbf{\LPBNN}: Approximating the posterior in a learned latent space of the rank1 vectors improved convergence, as LP-BNN outperformed Rank1 BNN. However, LP-BNN did not perform as well as other methods with regard to any metrics, likely because training layer-wise VAEs complicates the learning task. 

\textbf{\SWAG}: The SWAG models did not outperform the base models in terms of accuracy as expected. This can be attributed to the fact that the base models used in the evaluation were those resulting from the epoch with the best validation performance, whereas the SWAG weight posterior was fit across the converged SGD trajectory. This technique is desirable because it does not require adapting the architecture in any way and can be considered a post hoc process. 

\textbf{\MSWAG} Ensembling SWAG models improved the accuracy and UQ calibration but again at the expense of scalability. 

This benchmark provides some insights into how UQ methods can be improved. The Multi-SWAG performance reinforces the notion that ensembling Bayesian methods can improve the fidelity of approximate inference by enabling multimodal marginalization. This could be made more scalable by combining SWAG with a batch ensemble model rather than applying naive ensembling. Existing work has also demonstrated that combining ensembling with dropout improves performance on related medical imaging tasks \cite{adams2023fully}. These experiments additionally demonstrate that LP-BNN is a desirable alternative to Rank1 BNN. However, improvements could be made to the LP-BNN process of learning a low-dimension representation of the rank1 vectors, as layer-wise VAEs increase the training burden and hyperparameters to tune. 
The pancreas OOD analysis reveals that none of the epistemic UQ methods were effective at detecting instances with larger tumor sizes than those seen in training. As Table \ref{table:pancreas} demonstrate, all methods provided better calibrated uncertainty estimates on in domain test data than OOD.
Such failure has been noted before, as model misestimation can result in overconfidence in OOD predictions \cite{zhang2021understanding,besnier2021triggering}.
This illustrates the need to consider alternative test statistics and objectives in developing epistemic uncertainty estimation techniques. 

In conclusion, our benchmarking study of scalable epistemic uncertainty quantification techniques for challenging organ segmentation tasks highlights the importance of accurate uncertainty estimation in medical image analysis. The insights gained from this study can guide researchers and practitioners in selecting appropriate methods to enhance the reliability and robustness of deep learning models for organ segmentation, ultimately contributing to improved diagnosis and treatment planning in clinical practice.

\subsubsection{Acknowledgements}
This work was supported by the National Institutes of Health under grant numbers NIBIB-U24EB029011, NIAMS-R01AR076120, \\ NHLBI-R01HL135568, and  NIBIB-R01EB016701.
The content is solely the responsibility of the authors and does not necessarily represent the official views of the National Institutes of Health.
The authors would like to thank the University of Utah Division of Cardiovascular Medicine for providing left atrium MRI scans and segmentations from the Atrial Fibrillation projects.
%

%

\bibliographystyle{splncs04}
\bibliography{ref}

\end{document}